# Meteorite cloudy zone formation as a quantitative indicator of paleomagnetic field intensities and cooling rates on planetesimals

Clara Maurel, Benjamin P. Weiss, James F. J. Bryson

*a Department of Earth, Atmospheric and Planetary Sciences, Massachusetts Institute of Technology, Cambridge, MA, USA*

*b Department of Earth Sciences, University of Cambridge, Cambridge, UK*

## Abstract

Metallic microstructures in slowly-cooled iron-rich meteorites reflect the thermal and magnetic histories of their parent planetesimals. Of particular interest is the cloudy zone, a nanoscale intergrowth of Ni-rich islands within a Ni-poor matrix that forms below ~350°C by spinodal decomposition. The sizes of the islands have long been recognized as reflecting the low-temperature cooling rates of meteorite parent bodies. However, a model capable of providing quantitative cooling rate estimates from island sizes has been lacking. Moreover, these islands are also capable of preserving a record of the ambient magnetic field as they grew, but some of the key physical parameters required for recovering reliable paleointensity estimates from magnetic measurements of these islands have been poorly constrained. To address both of these issues, we present a numerical model of the structural and compositional evolution of the cloudy zone as a function of cooling rate and local composition. Our model produces island sizes that are consistent with present-day measured sizes. This model enables a substantial improvement in the calibration of paleointensity estimates and associated uncertainties. In particular, we can now accurately quantify the statistical uncertainty associated with the finite number of islands acquiring the magnetization and the uncertainty on their size at the time of the record. We use this new understanding to revisit paleointensities from previous pioneering paleomagnetic studies of cloudy zones. We show that these could have been overestimated by up to one order of magnitude but nevertheless still require substantial magnetic fields to have been present on their parent bodies. Our model also allows us to estimate absolute cooling rates for meteorites that cooled slower than < 10,000°C My$^{-1}$. We demonstrate how these cooling rate estimates can uniquely constrain the low-temperature thermal history of meteorite parent bodies. Using the main-group pallasites as an example, we show that our results are consistent with the previously-proposed unperturbed, conductive cooling at low temperature of a ~200-km radius main-group pallasite parent body.

**Keywords**

Planetesimals; Iron meteorites; Cloudy zone; Spinodal decomposition; Cooling rates; Extraterrestrial paleomagnetism

## 1. Introduction

Planetesimals, the ~1- to ~1000-km building blocks of planets, accreted within the first few million years (My) of the solar system (Hevey and Sanders, 2006). The existence of iron and stony-iron meteorites demonstrates that some of these planetesimals underwent large-scale melting and differentiation (McCoy et al., 2006). As these planetesimals cooled and solidified, their metal grains progressively formed different microstructures and minerals (Buchwald, 1975), whose existence and nature depend on the initial composition of the metal and its cooling rate. Understanding their formation can provide key constraints on the history of iron-rich meteorites and early accreted planetesimals.

The metal grains in iron meteorites, stony-iron meteorites and iron-rich chondrites are dominantly Fe-Ni in composition, alloyed with some minor elements (e.g., C, S, P, Cr, Si; Goldstein et al., 2009a). For bulk Ni contents between ~5.5 and ~19 wt.%, the Widmanstätten pattern develops within the Fe-Ni alloy as an intergrowth of Ni-poor $\alpha$-bcc (body centered cubic) kamacite and Ni-rich $\gamma$-fcc (face centered cubic) taenite during cooling between ~800°C and ~600°C, with the precise temperature range depending on the bulk Ni and P contents (Yang and Goldstein, 2005). During its formation, the growth of kamacite is controlled by temperature-dependent diffusion such that the width of kamacite lamellae strongly depends on the cooling rate of the meteorite. Below ~350°C, another phase separation occurs in the portion of the Ni-rich $\gamma$-fcc taenite phase located near the kamacite/taenite interface (Yang et al., 1996). This phase separation, called spinodal decomposition, results in the formation of the cloudy zone (CZ), a nanoscale intergrowth of ferromagnetic Ni-rich taenite crystals (known as islands) embedded in a Ni-poor, paramagnetic matrix with the structure of the fcc mineral antitaenite (Blukis et al., 2017). Like the size of the kamacite lamellae, the size of CZ islands is inversely related to the cooling rate.

For the past five decades, different techniques simulating the diffusion-controlled growth of the Widmanstätten pattern have been developed to determine the cooling rate of iron-rich meteorites (e.g., matching of the kamacite/taenite interface Ni profile or central Ni content; Goldstein et al., 2009a). Because this growth mostly occurs within ~100°C below the kamacite nucleation temperature (Goldstein and Ogilvie, 1964), these techniques provide an estimate of the meteorite's cooling rate between ~700°C and ~500°C. These cooling rates have significantly contributed to our understanding of the thermal evolution of meteorite parent planetesimals. For example, pioneering compositional measurements of iron meteorites combined with thermal modeling demonstrated that the parent bodies of all known iron meteorites were planetesimals rather than Moon-sized objects (Wood, 1964; Goldstein and Ogilvie, 1964). Cooling rate determinations also showed that planetesimals were fundamentally sculpted by catastrophic impacts (e.g., the IVA iron parent body that may have undergone one or several mantle-stripping impacts; Yang et al., 2008).

The correlation between size of the Ni-rich islands in the CZ and the cooling rate of meteorites has also long been identified as a potential cooling-rate indicator (Yang et al., 1997). Since the CZ forms below ~350°C, it would provide cooling rate estimates ~200°C below those recovered from the Widmanstätten pattern, providing additional constraints on late events not necessarily recorded by the Widmanstätten pattern like mild reheating or accretion of material (e.g., Goldstein et al., 2009b). Any prolonged reheating above ~350°C would result in the re-homogenization of the CZ region; if followed by an excavation, the incompatibility between the slow kinetics of spinodal decomposition (requiring cooling rates $\lesssim 10,000$°C My$^{-1}$) and the fast cooling ($\gtrsim 1,000,000$°C My$^{-1}$) of material exposed to space would prevent the CZ from reforming. In addition, the nm size of CZ islands (three orders of magnitude smaller than Widmanstätten structures) makes the CZ particularly sensitive to shock alteration (Goldstein et al., 2009a). The presence of the CZ is therefore indicative of a lack of reheating and shock events during the final cooling of the parent body. Despite this potential, a quantitative method has yet to be developed that provides an absolute estimate of the cooling rate at ~350°C from experimental measurements of CZ island size. Currently, island size measurements have only been used to determine the relative cooling rates of two meteorites and to relate the island size to the cooling rate at 700-500°C of a single meteorite using an empirical power-law (Yang et al., 2010).

The CZ also has the capability to preserve a record of the ambient magnetic field it experienced when it grew (Uehara et al., 2011). Such a record could be used to investigate whether a planetesimal generated a field by the dynamo process due to the advection of its molten metallic core (e.g., Bryson et al., 2015). This field-recording capacity is due to a phase transformation that occurs when the meteorite cools below 320°C at rates $\lesssim$ 5,000°C My$^{-1}$. At this temperature, the ferromagnetic $\gamma$-fcc taenite forming the Ni-rich CZ islands transforms into a tetragonal ferromagnetic mineral called tetrataenite ($\gamma''$). The fact that CZ islands are small (~15 to ~200 nm) and have the high magnetic coercivity associated with tetrataenite (> 1 T for the finest part of the CZ; Uehara et al., 2011) makes them exceptionally robust magnetic recorders.

It is particularly challenging to isolate the natural remanent magnetization (NRM) of CZ islands using traditional paleomagnetic techniques initially developed for analysis of mm- to cm-sized samples (Brecher and Albright, 1977). For example, one of the major impediments is the abundance of large ($\gg$100 μm; Buchwald, 1975) multidomain kamacite grains, which can be easily remagnetized (Dunlop and Özdemir, 1997) and could constitute the main source of the magnetic signal when measuring an iron meteorite sample. An alternative was recently developed to isolate the NRM carried by tetrataenite CZ islands (Bryson et al., 2014a). Using X-ray photoemission electron microscopy (XPEEM), the magnetization of the CZ alone can be measured at the nm-scale along several kamacite/taenite interfaces and used to calculate the relative orientation and the intensity of the ambient magnetic field present when the CZ grew.

Blukis et al. (2017) posed four fundamental questions that should be addressed in order to obtain more accurate paleointensity estimates from XPEEM images of the CZ: 1) What is the magnetic state of islands when they form? 2) What is their blocking temperature and how is their remanence changed when cooling through this temperature? 3) What is their volume at blocking temperature? 4) What is the influence of magnetostatic interaction between islands? The authors addressed question 1) by showing that the matrix phase of the cloudy zone is paramagnetic, implying islands can be seen as an ensemble of interacting grains of taenite above 320°C, and of tetrataenite below this temperature. Einsle et al. (2018) addressed question 2) both experimentally and with micromagnetic simulations, in which they assumed the whole crystallographic structure of an island readily orders at tetrataenite formation temperature (320°C). In this case, they showed that any NRM acquired by the parent taenite is lost during the taenite/tetrataenite phase transition and that an independent remanence is recorded—implying that the blocking temperature of CZ islands is 320°C, and that the CZ cannot provide a time-resolved record of the ancient magnetic field, as first suggested in pioneer XPEEM studies (e.g., Bryson et al. 2015). Question 4) is an area of active research. Currently, no interactions between islands are included in the equation used to estimate a paleointensity from XPEEM measurements. CZ islands are assumed to be an ensemble of single-domain "grains" with the orientation of their magnetic moment following a Maxwell-Boltzmann distribution (Bryson et al., 2014b). Interactions could affect the absolute paleointensity we estimate from one CZ, but it is very unlikely they could produce a uniform remanence over two separated CZ and lead to the false conclusion that a field was present when there was no field.

The present study addresses the issue of the volume of the islands (question 3), with two important implications for the estimation of ancient field intensities from XPEEM data. First, in the current Maxwell-Boltzmann framework, absolute paleointensity estimates are inversely related to the volume of the islands when they recorded a field at 320°C (see Supplementary Material of Bryson et al., 2017). Second, one important source of uncertainty on paleointensity estimates from the CZ comes from whether the net moment of the islands included in an XPEEM dataset is statistically representative of the ancient field. Berndt et al. (2016) showed that this statistical uncertainty is particularly sensitive to the island size at blocking temperature. The authors estimated that when CZ islands cooled through 320°C, their diameter was ~8 nm, implying that an impractically large number of ~$10^9$ islands should be sampled in each XPEEM dataset to obtain statistically meaningful paleodirections and intensities ($10^3$–$10^4$ islands are typically analyzed during a XPEEM experiment). This led Berndt et al. (2016) to question the reliability of published paleomagnetic XPEEM data. However, they obtained this estimate assuming CZ islands formed through nucleation and growth, a process different from spinodal decomposition.

Motivated by the implications of better understanding cloudy zone formation for low-temperature cooling rate determination and paleointensity estimation, we developed a one-dimensional (1D) numerical model of CZ formation by spinodal decomposition in the cooling environment of a meteorite parent body. For a given local Ni content, the model estimates the average CZ island equivalent diameter (hereafter island size) at any temperature as a function of cooling rate. It therefore provides 1) an absolute cooling rate estimate at ~350°C, thereby offering a new approach for studying the low-temperature thermal history of cloudy-zone-bearing meteorites, and 2) an accurate value for the size of the islands at blocking temperature, which is an important step toward the goal of estimating absolute paleointensities from XPEEM data.

## 2. Cloudy zone formation model

**2.1. Spinodal decomposition**

The coexistence of two (or more) phases at equilibrium can occur for a bulk composition lying within the miscibility gap on its phase diagram, where it is more energetically favorable for a homogeneous system to separate into these phases (Porter et al. 2009). The compositions that delimit the miscibility gap for a given temperature are those where the free energy curve possesses a common tangent. Between these two compositions, the free energy curve also possesses two points of inflection, characterized by a change in sign of the free energy's second derivative. These points separate the metastable region of the miscibility gap from its unstable region. The distinction between metastable and unstable regions is therefore related to the convex and concave shape of the Gibbs free energy curve, respectively. Consider a binary system, say an Y-Z alloy, with a bulk composition falling on the convex part of the curve (Fig. 1A–B). Small thermal fluctuations in composition (i.e., departure from the bulk composition toward Y-rich and Z-rich compositions, following the free energy curve) will necessarily increase the free energy of the system, making the separation into two phases energetically unfavorable (Fig. 1C top); such a system is metastable. For the phase separation to occur, this energy barrier will have to be overcome: this is the process of nucleation. Now, consider the composition of the Y-Z system lying on the concave part of the free energy curve (Fig. 1B, C bottom). Any infinitesimal, thermally-induced fluctuations in composition (inherent to any system) will necessarily decrease the free energy of the system and therefore *spontaneously* cause phase separation; this is the mechanism of spinodal decomposition.

Because it does not require any energy barrier to proceed, spinodal decomposition simply relies on diffusion of atoms in the two forming phases and is therefore governed by Fick's first law of diffusion:

$$J = -M\nabla\mu \qquad (1)$$

In this equation, $J$ is the diffusion flux (m$^2$ s$^{-1}$), $\mu$ is the chemical potential (kg m$^{-1}$ s$^{-2}$) and $M$, called atomic mobility (s kg$^{-1}$), is positive and proportional to the diffusion coefficients of each element in the alloy (e.g., Y and Z). This equation is a generalized expression for non-ideal solutions (i.e., with uneven interatomic forces) of the common form $J = -D\nabla X$ where $X$ is the concentration and $D$ is a diffusion coefficient. Cahn (1965) derived an expression of the chemical potential $\mu$ as a function of the composition and the Gibbs free energy density ($g$) of the system:

$$\mu = \frac{\partial g}{\partial X} - \nabla \cdot (2\kappa \nabla X) \tag{2}$$

where $\kappa$, called the gradient-energy coefficient, reflects the contribution of the local composition to the total energy of the system. Given that:

$$\nabla \cdot J = -\partial X/\partial t \tag{3}$$

one can re-write eq. (1) using eq. (2) and (3) to obtain the so-called Cahn-Hilliard equation of diffusion:

$$\frac{\partial X}{\partial t} = \nabla \cdot \left( M \frac{\partial^2 g}{\partial X^2} \nabla X \right) - \nabla \cdot \{M \nabla [\nabla \cdot (2\kappa \nabla X)]\} \tag{4}$$

Solving eq. (4) for $X$ provides dependences of the composition on space (i.e., the size of the CZ islands) and time (or equivalently temperature). Both dependences must be known to use our model to 1) calculate the statistical uncertainty of CZ paleomagnetic measurements, 2) estimate an absolute field intensity, and 3) estimate an absolute cooling rate at ~350°C.

We can analyze eq. (4) to understand the various stages of spinodal decomposition. Any system is subject to local thermally-induced fluctuations in composition. These fluctuations can be expressed as a sum of spatial sinusoids with characteristic wavelengths. Spinodal decomposition leads to the selective amplification of some of these wavelengths. Let us first take an example where an ideal system is instantaneously quenched to and kept at a temperature within the spinodal region. Making the simplifying assumption that $M$, $\kappa$ and $\frac{\partial^2 g}{\partial X^2}$ are independent of the composition $X$ (i.e., $g$ is a cubic polynomial function of $X$, Fig. 1D), one can solve analytically eq. (4) for $X$ and find a solution in the form of a Fourier series (Hilliard, 1970), which describes how quickly the growth of fluctuations at a given wavelength will be. If only the first term of the right-hand side of eq. (4) is taken into account, the solution yields infinitesimally small wavelengths infinitely amplified. In reality, the second term of the right-hand size of eq. (4), related to the energy cost of an interface (via $\kappa$), prevents very small wavelengths from growing to limit the creation of interfacial area and associated excess of energy. This balance between the two right-hand side terms of eq. (4) results in the existence of a preferred wavelength that receives the maximum amplification (Hilliard, 1970). For reference, this wavelength ($\lambda_{\text{pref}}$) is given by:

$$\lambda_{\text{pref}}^2 = -\frac{16\pi^2 \kappa}{\frac{\partial^2 g}{\partial X^2}} \quad \text{for} \quad \frac{\partial^2 g}{\partial X^2} < 0 \tag{5}$$

Once the system is ideally quenched within the spinodal, the size of proto-islands will peak around the preferred wavelength, forming a relatively periodic two-phase pattern that, recalling the example of the Y-Z system above, is composed of alternating Z-rich and Z-poor phases (islands and matrix, respectively). Accounting for the dependence of $g$ on $X$ in real systems (as opposed to the cubic approximation adopted above) the Z content of the two phases will evolve toward the local minima of the free energy curve (Fig. 1D), which corresponds to the miscibility gap boundaries (Fig. 1A). Meanwhile, the excess in surface energy at the interfaces between islands and matrix resulting from the sharp concentration gradients will gradually become the dominant force in the system (Cahn, 1966). To reduce their surface energy, islands will start a much slower coarsening process where large islands may grow at the expense of smaller ones.

Unlike this conceptual example of a quenched alloy, a meteoritic alloy will in reality slowly cool through the spinodal boundary and continue cooling after spinodal decomposition has started. However, the overall behavior of the system is similar to that described above, with the exception that both the preferred wavelength and the amplification factor vary with temperature (Hutson et al., 1966). According to eq. (5), at the onset of spinodal decomposition (where $\frac{\partial^2 g}{\partial X^2} = 0$) the preferred wavelength is theoretically infinite. However, within less than a degree below the spinodal temperature (where $\frac{\partial^2 g}{\partial X^2} < 0$), the preferred wavelength has decreased exponentially and fluctuations of the order of tens of nm start to grow (see Cahn, 1968). The fact that the preferred wavelength decreases with temperature simply results in a broadened size distribution of the fluctuations (i.e., the islands) because different wavelengths will be favored as spinodal decomposition progresses. Finally, the coarsening rate will decrease with temperature due to the slower diffusion rate (see Fig. S1.1). Our model solves eq. (4) for $X$ to obtain the CZ island size as a function of temperature $T$. However, to solve the equation, one must first find the dependences of $g$, $\kappa$ and $M$ on $X$ and $T$. The dependences on temperature and composition of $g$ are summarized in the following section. A similar analysis for $\kappa$ and $M$ is made in Supplementary Material S1.

**2.2. Gibbs free energy density, $g$**

The free energy density of an alloy depends on both its composition and temperature. For Fe-Ni, spinodal decomposition occurs in the $\gamma$-fcc phase with both islands and matrix remaining as $\gamma$ phases for most of their growth time. As a consequence, we do not account for a variation in energy due to a modification of the crystal structure (Section 6). Cacciamani et al. (2010) derived an analytical expression for the Gibbs free energy for Fe-Ni using experimental data available coupled with atomistic calculations. The free energy density of a given phase is the sum of four contributions:

$$g(X,T) = g_{\text{ref}}(X,T) + g_{\text{id}}(X,T) + g_{\text{ex}}(X,T) + g_{\text{mag}}(X,T) \qquad (6)$$

where $g_{\text{ref}}$ is the reference free energy density of the pure elements, $g_{\text{id}}$ is the free energy density for ideal mixing (that of an equivalent ideal mixture), $g_{\text{ex}}$ is the excess free energy density (accounting for the non-ideality of the system) and $g_{\text{mag}}$ is the magnetic contribution. The description of each term is given in Supplementary Material (S1).

Using eq. (6), we can calculate the free energy density of $\gamma$-fcc Fe-Ni as a function of composition for a given temperature (Fig. 2A). Though initially subtle, the region of spinodal decomposition (which lies between the two inflection points) becomes more readily visible with decreasing temperature. In the case of $\gamma$-Fe-Ni, the shape of the free energy curve and therefore the existence of the spinodal region is influenced by the higher-order phase transition from paramagnetic to ferromagnetic (accounted for in the term $g_{\text{mag}}$). As a consequence, if spinodal decomposition occurs before any other phase transition (e.g., taenite ordering into tetrataenite), the two phases involved in spinodal decomposition have the same crystal structure and only differ by their magnetic properties.

The points of inflection and the points of common tangent derived from eq. (6) determine the boundaries on the Fe-Ni phase diagram of the spinodal region and miscibility gap, respectively (Fig. 2B). As noted by Cacciamani et al. (2010), the spinodal boundaries obtained with this analytical expression slightly differ from the Fe-Ni metastable phase diagram proposed by Yang et al. (1996), which serves as a reference among the meteorite community. Cacciamani et al. (2010) still concluded that given the uncertainties arising from the challenging experimental identification of metastable equilibria, these two spinodal boundaries are in good agreement with each other. Note that the spinodal boundaries proposed in Yang et al. (1996) were based on the observation by Reuter et al. (1989b) of a correlation between the presence of ordered tetrataenite and the presence of a spinodal decomposition product. Yang et al. (1996) assumed that the system entered the spinodal region at the same temperature as that of tetrataenite ordering, but there is no evidence in their experimental data that spinodal decomposition did not occur at a higher temperature.

### 2.3. Model implementation

Once the dependence on composition $X$ and temperature $T$ of $g$, $\kappa$ and $M$ in eq. (4) are specified, we can solve eq. (4) for $X$ as a function of space and temperature. For this, we use the Python package Fipy, a partial differential equations solver based on the finite volume method (Guyer et al., 2009). We investigate bulk compositions and cooling rates ranging from 35 to 41 wt.% Ni and 1 to 10,000°C My$^{-1}$, respectively. The system starts at the temperature defined by the spinodal boundaries at the given bulk composition. To simulate a cooling environment, we decrease the temperature by steps of 0.1°C assuming a linear relationship between time and temperature; the expression of each variable ($M$, $\kappa$ and $g$) is updated after such temperature step. More details can be found in Supplementary Material S1.

# 3. Results

## 3.1. Evolution of the island size and island/matrix composition

For a given Ni composition and cooling rate, we simulated the growth of CZ islands in a cooling environment from the temperature dictated by the spinodal boundary (Fig. 2B) down to 210°C, when the island size stops growing due to the extremely slow diffusion (Fig. 3). Within < 1°C of crossing the spinodal, the initially infinitesimal fluctuations in composition begin to amplify. After typically a few tenths of a degree, the islands reach their equilibrium composition (~45 wt.% Ni, Fig. 4A), which we find to be the same regardless of bulk composition and cooling rate of the system. This composition is ~2-3 wt.% less than the measured Ni concentration of the CZ islands (Goldstein et al., 2009a; Einsle et al., 2018) and than that obtained analytically with Eq. (6) (Fig. 2B). This small difference is likely due to one or more approximations used in the model implementation (e.g., the approximation of the free energy curve—see Supplementary Material S1.4). When the islands reach their equilibrium composition, their size is between ~40% and ~55% of their present-day size depending on cooling rate and bulk Ni content. Subsequently, the islands slowly grow by coarsening, resulting in a decrease in the matrix Ni content to keep the bulk composition of the system constant (Fig. 4A–B). By the time the system reaches 210°C, the diffusion rate has dropped by ten orders of magnitude (see Fig. S1.1) and both island size and matrix compositions become stationary. For a given cooling rate, the final Ni contents in the matrix for bulk compositions between 35 and 41 wt.% Ni vary by ~1.5 wt.%. The variation in matrix composition with cooling rate for a given bulk Ni content is more pronounced, ranging for example from ~24 wt.% Ni at 5000°C My$^{-1}$ to ~17 wt.% Ni at 5°C My$^{-1}$ for a 40 wt.% Ni alloy.

### 3.2. Island size at 320°C

One aim of our model is to improve the reliability of paleointensities recovered from XPEEM data of the CZ. Based on the diffusion length of Ni in taenite, previous paleomagnetic studies that provided absolute paleointensity estimates (Bryson et al., 2015; Nichols et al., 2016) have assumed that the CZ islands in slowly cooled meteorites (<100°C My$^{-1}$) were 30% of their present-day size when they recorded a field. Here, we find that the island size at 320°C is almost independent of the cooling rate and ranges between 60% and 85% of present-day size for bulk Ni contents between 35 and 41 wt. %, respectively (Fig. 5). Additionally, in the Maxwell-Boltzmann framework currently used to estimate a paleointensity from XPEEM data, the intensity is inversely proportional to the volume of the islands (supplementary material of Bryson et al., 2017); using 30% instead of 60–85% of present-day size therefore results in a likely overestimation of the paleointensity by a factor of ~8–20.

### 3.3. Present-day island size

By the end of the simulations, the island size is essentially equivalent to the present-day size. The CZ island size inversely correlates with the cooling rate and the bulk Ni content, which decreases with distance to the tetrataenite rim (e.g., Goldstein et al., 2009a). Both correlations are reproduced in our results (Fig. 6). Most islands with measured sizes are those next to the tetrataenite rim—the largest in the CZ. If a meteorite cooled very slowly (≲ 5-10°C My$^{-1}$), this region will have a composition of ~40-42 wt.% Ni according to the equilibrium boundaries of the Fe-Ni phase diagram. For these conditions, our model predicts island sizes >100 nm, which is in good agreement with islands sizes measured in the pallasites and mesosiderites, which are slowly cooled meteorite groups (Yang et al., 2010; Hopfe and Goldstein, 2001). The other lines on Fig. 6 are applicable to faster-cooled meteorites (e.g., IVA, IVB, IIIAB), for which the bulk composition next to the rim can be lower than 40 wt.% Ni because equilibrium could not be reached (e.g., Goldstein et al. 2009b). The decrease in island size with distance to the rim (due to the decrease in Ni content) can be seen as a vertical line at a given cooling rate down Fig. 6. Note that the model assumes a constant local composition, which corresponds to a narrow band parallel to the tetrataenite rim. Therefore, we can currently only model stepwise decreases in island size with distance to the rim (by using different initial compositions). A continuous decrease of the Ni content with distance to the rim will be the object of future improvements to the code.

To test our model against experimental data, three pieces of information are needed: the average island size in a given region, a high-resolution Ni composition of this region, and an independent cooling rate estimate at ~350°C. The H6 chondrite Guareña has all three pieces of information essentially available (with the caveat that the composition profile has a coarse resolution of 1 μm). An approximate cooling rate of ~3.7°C $My^{-1}$ between ~450 and ~250°C can be deduced from the difference between Guareña's U-Pb age determined on phosphates and Ar-Ar age determined on feldspars given the closure temperatures of these thermochronological systems (Henke et al., 2013). Guareña has an average island size of 120 ± 5 nm and the composition profile shows a composition between 41 and 39 wt.% Ni next to the tetrataenite rim (Scott et al., 2014). For this composition, our model predicts a cooling rate at ~350°C of 4.4 ± 3.2°C $My^{-1}$, in agreement with the aforementioned value.

## 4. Cooling rate application: example of the main-group pallasites

Our CZ formation model is a promising tool for investigating the cooling history of iron meteorite and iron-rich chondrite parent bodies at a temperature range that has been previously poorly constrained. Combined with cooling rate estimates at 700-500°C, these data can place new constraints on long-term planetary thermal evolution. However, measurements of the bulk composition near the tetrataenite rim and CZ island size are necessary to fully take advantage of the model and it is rare that both of these properties have been measured and published. It is beyond the scope of this paper to present such data. Nevertheless, as an example of application of our model as cooling rate indicator, we consider the case of the main-group pallasites, for which island sizes, some low-resolution composition profiles, and a parent-body thermal model have been published.

Yang et al. (2010) measured the island size of seven very slow-cooled pallasites and determined cooling rates at 700-500 °C from Widmanstätten taenite profile-matching, ranging between $2.5 \pm 0.3$°C My$^{-1}$ and $8.9 \pm 1.2$°C My$^{-1}$. The authors presented composition profiles for the Giroux pallasite with 1-µm resolution, finding a bulk Ni content of ~40 wt.% next to the rim. Using a bulk Ni content between 39 and 40 wt.%, the model predicts cooling rates at ~350°C ranging between $0.7 \pm 0.4$°C My$^{-1}$ and $3.2 \pm 1.8$°C My$^{-1}$ (Fig. 6). Our calculated cooling rates at ~350°C indicate that these meteorites cooled ~0.35 times slower at ~350°C than at 700-500°C (Fig. 7). Using thermal evolution models, Bryson et al. (2015) proposed that the main-group pallasite parent body was a fully differentiated body with a diameter of ~400-km. According to their model, the cooling rate at ~350°C for the pallasites considered in this study should be ~0.5 times the cooling rate at 700-500°C. Our results therefore support the idea that these pallasites cooled without undergoing any significant reheating or shock event mantle of their ~400-km parent and not deeper at the core-mantle boundary (Yang et al., 2010; Tarduno et al., 2012; Bryson et al., 2015).

## 5. Paleomagnetic application

Berndt et al. (2016) derived in the Maxwell-Boltzmann framework—also currently assumed for XPEEM—the number of magnetic carriers necessary to obtain a given uncertainty in paleodirection and paleointensity due to the limited number of islands included in a dataset (called statistical uncertainty in the following; Supplementary Material S2). This number inversely depends on the CZ island size at blocking temperature, which our model now constrains (Fig. 5). Intuitively, this is because larger islands have larger magnetic moments and so couple more strongly to an external field such that fewer islands are required to achieve the same net magnetization. The required number of islands also inversely depends on the intensity of the ancient field—more islands are needed for their net moment to be representative of a weak field. Therefore, with a reasonable assumption about the intensity of the ancient field, one can combine the results of our model with the formula of Berndt et al. (2016) to calculate how many islands will be needed in a future XPEEM dataset to limit the statistical uncertainty in paleodirection and paleointensity to a given value (Fig. 8A).

Similarly, knowing the number of islands included in a XPEEM dataset and with a paleointensity estimate, one can quantify the statistical uncertainty for published datasets. The model also allows us to quantify an uncertainty on island size at 320°C. Adding both types of uncertainty to the measurement uncertainty allows us to more accurately represent the total uncertainty of the paleointensity estimate. In light of these results, we review previously published XPEEM studies for which a paleointensity estimate has been proposed (Fig. 9, Table 1).

## 5.1. Main-group pallasites: Imilac and Esquel

The first meteorite paleomagnetic study using the XPEEM technique was conducted by Bryson et al. (2015) on the Imilac and Esquel main-group pallasites. These meteorites possess cloudy zones with ~143-nm and ~157-nm diameter islands near the tetrataenite rim, respectively

(Yang et al., 2010). The authors measured the magnetization of four non-overlapping 4500×400-nm regions next to the tetrataenite rim for both meteorites. Assuming the islands occupy 90% of a region's area, a total of ~320 islands (Imilac) and ~380 islands (Esquel) was included in each dataset. Adopting an island size at 320°C equal to 30% of present-day size, Bryson et al. (2015) estimated paleointensities of 120 µT for Imilac and 84 µT for Esquel with a $2\sigma$ uncertainty due to measurement noise of 10% and 16%, respectively. Using present-day island sizes, they obtain average paleointensities of 3.2 µT for Imilac and 2.2 µT for Esquel. It should be noted that independently of the island size adopted, these values are lower limits on the paleointensity estimate because the sample was only measured in one orientation (i.e., only a combination of the three components of the paleofield was calculated; see supplementary material of Bryson et al., 2017).

Assuming a bulk composition between 40 and 39 wt.% (i.e., CZ islands at 320°C about 78% of their present-day size; Fig. 5), the average paleointensities become 6.8 µT (Imilac) and 4.8 µT (Esquel). Using these estimates, we calculate a statistical uncertainty ($2\sigma$) in paleointensity of 23% for both Imilac and Esquel (Fig. 8B). In addition, a $2\sigma$ uncertainty of ±5 % for the island size at 320°C (equivalent to a ±1 wt.% Ni uncertainty in composition) would result in a 15% uncertainty on paleodirection/paleointensity. After combining these uncertainties, the paleointensities become 6.8 ± 2.0 µT and 4.8 ± 1.5 µT.

Using data provided by Bryson et al. (2015), we simulated an XPEEM dataset that would have been measured if the CZ had cooled through 320°C in the absence of a magnetic field. We calculated the field intensity resulting from this dataset and proceeded by bootstrapping to obtain the upper bound of the 95 % confidence interval of this "zero-field" intensity, equal to 1.7 µT for Imilac and 1.1 µT for Esquel (procedure described in the supplementary material of Bryson et al.,

2017). Although the paleointensities are revised downward, they are both larger than these values and therefore still require a substantial magnetic field on the parent body, indicating the past existence of a core dynamo.

The improved paleointensities now differ from the paleointensities estimated by Tarduno et al. (2012) using olivine grains on Imilac (72.7 ± 7.1 µT) and Esquel (125.2 ± 12.9 µT), especially given that the olivine grains may have been shielded from the planetary field by swathing kamacite (see supplementary information for Bryson et al. 2015). Given the different blocking temperatures for taenite in olivine grains (~550°C) and tetrataenite in the metal phase (320°C), one possible explanation for this apparent discrepancy is that silicates and metal simply recorded the magnetic field at different times of its history. It should also be noted that in the absence of a model describing the possible magnetostatic interactions between islands, our revised intensities—like other intensities discussed below—should be considered as more accurate but not final values.

**5.2. Main-group pallasites: Brenham and Marjalahti**

Nichols et al. (2016) used XPEEM to study the Brenham and Marjalahti pallasites (~123-nm and ~118-nm islands, respectively; Yang et al., 2010). The authors analyzed twelve and nine 4500 × 450-nm regions resulting in 1800 islands and 1480 for Brenham and Marjalahti, respectively. Average intensities of 4 µT for Brenham and 5 µT for Marjalahti were reported (without measurement uncertainties). With islands ~78% of their present-day size at 320°C (i.e., local Ni content of 39.5 wt.%), the average intensities become 0.2 and 0.3 µT. Like for Imilac and Esquel, we calculated the upper bound of the 95-% confidence of a "zero-field" intensity and found 1.1 µT and 1.4 µT for Brenham and Marjalahti, respectively. The fact that these values are larger than the paleointensity estimates above agrees with the conclusion by Nichols et al. (2016) that we cannot reject the hypothesis that Brenham and Marjalahti cooled through 320°C in the absence of a field. This is therefore consistent with the liquid core of the pallasite parent body experiencing a quiescent period before its period of compositional convection induced by crystallization.

### 5.3. IVA iron: Steinbach

Bryson et al. (2017) applied the XPEEM technique to the IVA iron Steinbach. The magnetization was measured along two CZ, imaging nine 4500×100-nm regions along each. Adjacent to the tetrataenite rim, Steinbach's islands are 29 nm in diameter, such that ~5,500 islands were included in each dataset. The authors reported a paleointensity of ~100 µT (using present-day island size) with a ~50% measurement uncertainty.

No bulk composition profile has been published for Steinbach. However, such a profile was measured by Goldstein et al. (2009b) for the Chinautla IVA iron, which has a similar cooling rate at 700-500°C (~110°C My$^{-1}$) as Steinbach (~150°C My$^{-1}$). The average composition of the region with the coarsest islands in Chinautla is 37.5 wt.% Ni. With this composition CZ islands were ~70% of their present-day size at 320°C, and the average paleointensity becomes ~290 µT. With such small islands but intense field, the statistical uncertainty is 22%—the fact that it is very similar to that calculated for Imilac and Esquel is fortuitous (Fig. 8B). Combining this with the ~50% measurement uncertainty and the 15% uncertainty accounting for uncertainty in Ni content, we obtain a paleointensity estimate of 290 ± 165 µT. Bryson et al. (2017) concluded that the IVA parent body generated a field, strong and directionally-varying. Our results suggest that the field intensity has a large total uncertainty but do not invalidate the conclusion that a non-zero directionally-varying field was present.

## 6. Discussion: Effect of the taenite to tetrataenite transition

Our model is based on the free energy equations for the fcc $\gamma$-Fe-Ni phase and we do not consider any possible phase transitions that occur during the growth of the CZ. In particular, we do not model the ordering from taenite to tetrataenite at 320°C. This transformation will likely change the free energy curves (Fig. 2A) and possibly affect the growth of CZ islands, but the contribution of this phase to the total free energy is essentially unknown given that the cooling rates required for tetrataenite formation are unachievable on laboratory timescales. Hence, our model is effectively only valid for the regions of the CZ where spinodal decomposition started at temperatures above 320°C. According to the metastable Fe-Ni phase diagram (Fig. 2B), this corresponds to compositions $\gtrsim$ 34 wt.% Ni. The exact consequences of tetrataenite formation on the growth of already large islands are unclear. We can only speculate that, given the good agreement between our model and experimental data (Sections 3.3 and 4), the effect of the phase transformation for regions of the CZ above 35 wt.% Ni may only be minor.

In most meteorites studied with XPEEM, a clear difference has been observed between the XMCD signal of the coarse-to-medium CZ and the fine CZ. The latter shows a strong dominance of one easy axis, as opposed to the former, where the bias may be present but less clearly visible. Einsle et al. (2018) proposed two explanations: 1) these fine islands were single-domain taenite grains above tetrataenite formation temperature, as opposed to pseudo-single domain (vortices) coarse-to-medium islands, and interacted more strongly, or 2) spinodal decomposition occurred below the tetrataenite formation temperature in the fine CZ. Based on the updated Fe-Ni phase diagram (Fig. 2B), we could certainly discriminate between these two hypotheses in a future study by measuring the local Ni content at high resolution across the CZ and compare it with the location of the change in magnetic behavior between the medium and the fine CZ.

To study the effect of tetrataenite transition on an island's magnetization, Einsle et al. (2018) modeled entire CZ islands changing their crystallographic structure at 320°C. It is, however, conceivable that tetrataenite may not have ordered all at once in an island, implying that the new NRM may be acquired over a broader range of temperatures. This question remains unanswered and is the object of active research. It is beyond the scope of this study to speculate on the effect of a gradual transformation. It can simply be said that our model could as well provide the island size at any other "blocking" temperature.

## 7. Conclusion

- We developed a numerical model of cloudy zone (CZ) formation by spinodal decomposition in the cooling environment of a meteorite parent body.
- This model provides the compositional and structural evolution of the CZ islands as well as the size of the islands at 320°C, when they could record an ambient magnetic field.
- This island size allows us to quantify the uncertainty on paleodirection and paleointensity due to the limited number of magnetic carriers in experimental datasets. Combined with the uncertainty in island size at blocking temperature and measurement uncertainty, this provides a more accurate total uncertainty of the estimates.
- The model allows us to determine more accurately the intensity of a putative ancient field recorded by the CZ. Current research aims at understanding how the magnetostatic interactions between islands might affect the absolute paleointensities.
- This model also serves as an absolute cooling rate indicator that can provide new constraints on the low-temperature history of iron meteorite and iron-rich chondrite parent bodies.

# Acknowledgements

We gratefully thank Dr. Bradley R. Nakanishi (MIT) for helpful discussions about metallic alloys and phase transformations, as well as Dr. Trevor Keller (NIST) and the Fipy community for assistance with the Fipy package. We also thank Dr. Claire I. O. Nichols and Dr. Richard J. Harrison (University of Cambridge) for constructive discussions. C. M. and B. P. W. thank Thomas F. Peterson, Jr. and the NASA Discovery program for support. J. F. J. B. would like to thank St. John's College, University of Cambridge for financial support. We thank Dr. Thomas Berndt, Dr. Minoru Uehara and an anonymous reviewer for their helpful reviews.

# Figures

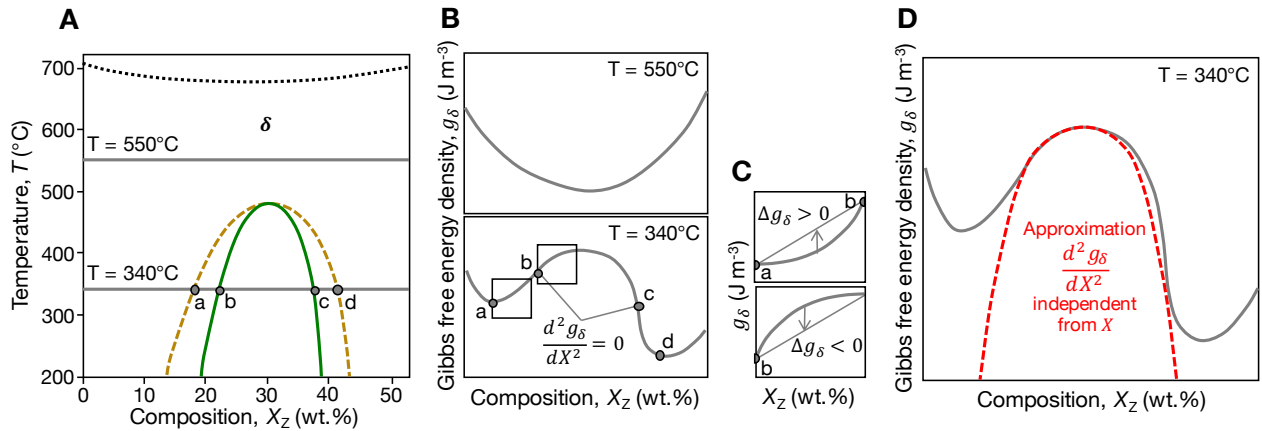

Fig. 1. A) Schematic of a low-temperature phase diagram for a hypothetical Y-Z compositional system. The black dotted line represents the equilibrium boundary for the phase $\delta$. The tan dashed line represents the miscibility gap boundaries and the full green line represents the spinodal boundaries. Full horizontal lines show temperatures at which Gibbs free energy density is described in (B) and (C). B) Schematic of the Gibbs free energy density $g$ of the phase $\delta$ as a function of the content in element Z, at a temperature of 550°C (top) and 340°C (bottom). Points "a" and "d" correspond to the point of common tangent that dictates the miscibility gap boundaries. Points "b" and "c" show the points of inflection, which determine the spinodal boundaries. C) Sketch of the effect of inherent fluctuations in composition. In the convex part of the free energy curve (top), any fluctuation around a given mean composition tends to increase the free energy (up arrow): the growth of these fluctuations is not energetically favorable and the system is metastable. In the concave part of the free energy curve (bottom) even the smallest fluctuation yields a decrease in energy (down arrow): the growth of the fluctuations is in that case favored and spontaneous; the system is unstable and spinodal decomposition occurs. D) Schematic of the approximation of the Gibbs free energy curve historically employed to solve analytically the Cahn-Hilliard equation (Hilliard, 1970).

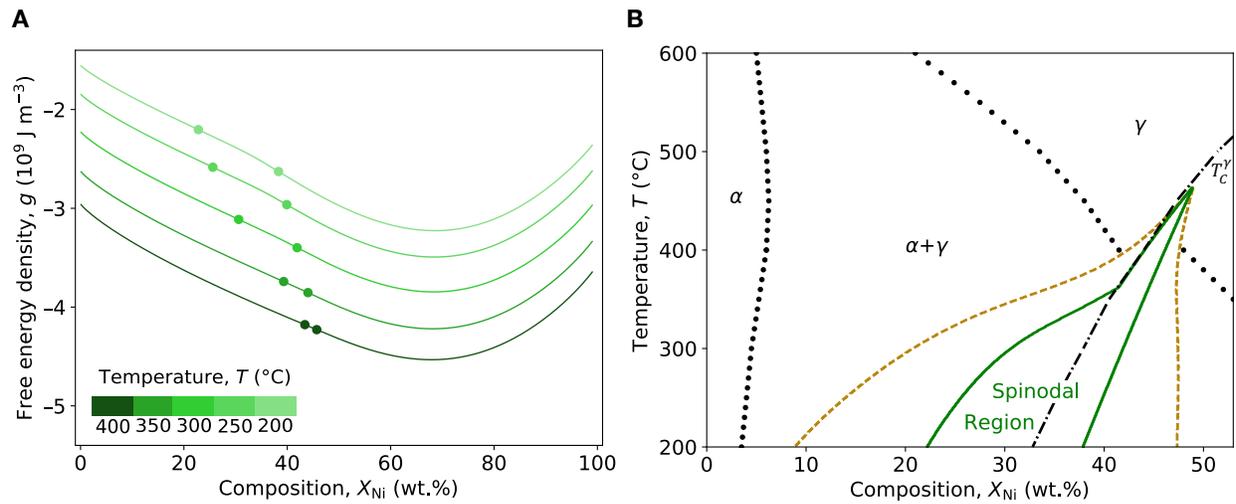

Fig. 2. A) Gibbs free energy density as a function of Ni composition for temperatures between 400°C and 200°C obtained from eq. (6). Colors denote the temperature at which they are calculated. Dots highlight the location of the points of inflection. B) Low-temperature phase diagram for the Fe-Ni system obtained from eq. (6). Stable phase equilibria for the $\alpha$ (kamacite) phase and $\gamma$ (taenite) phase are shown by the black dots. Green full and tan dotted lines represent the spinodal boundaries and the metastable phase equilibria (miscibility gap), respectively. The dash-dot line shows the Curie temperature of $\gamma$-fcc ($T_C^{\gamma}$) as a function of Ni content (from Cacciamani et al. 2010).

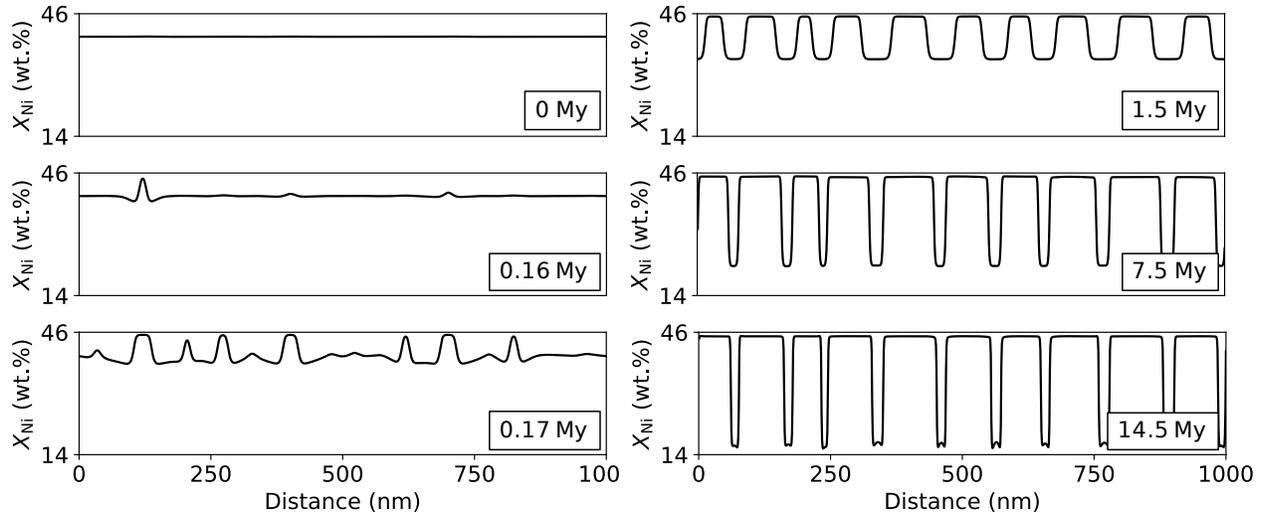

Fig. 3. Formation of the cloudy zone as simulated with our one-dimensional (1D) numerical model. Shown is the Ni content as a function of distance within the cloudy zone at six different steps of the phase separation. This 1D section can be seen as a band parallel to the tetrataenite rim of constant local Ni composition. The initial bulk composition of the system is 40 wt.% Ni and the cooling rate is 10°C My$^{-1}$. Times (lower left of each frame) are relative to the time the system cools through the spinodal boundary on the Fe-Ni phase diagram (Fig. 2B).

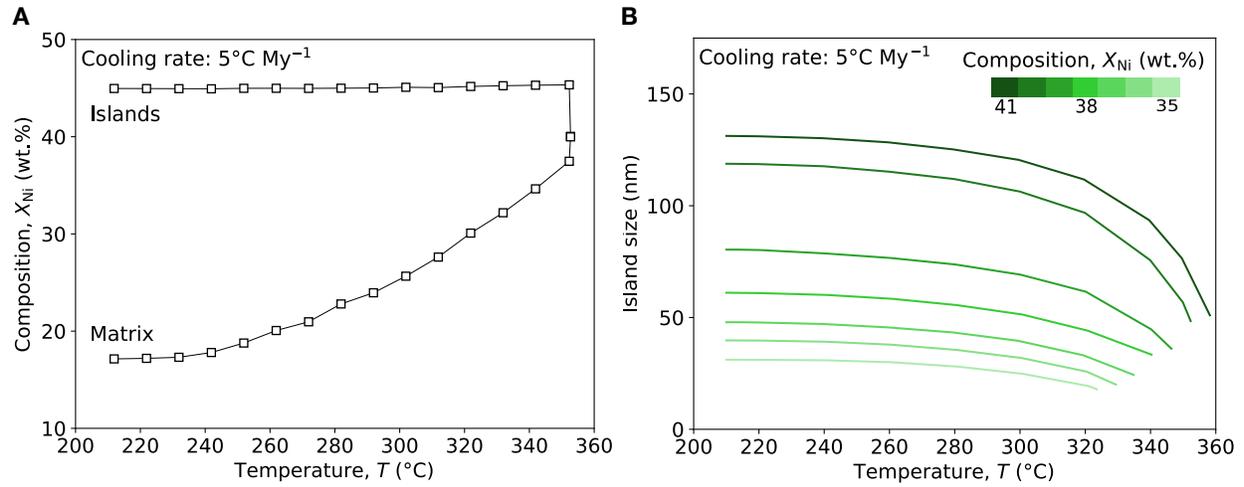

Fig. 4. A) Composition of the islands (upper curve) and the matrix (lower curve) for a local bulk composition of 40 wt.% Ni as a function of temperature for a cooling rate of 5°C My$^{-1}$. B) Average island size as a function of temperature for a cooling rate of 5°C My$^{-1}$ and bulk compositions of 35 to 41 wt.% Ni.

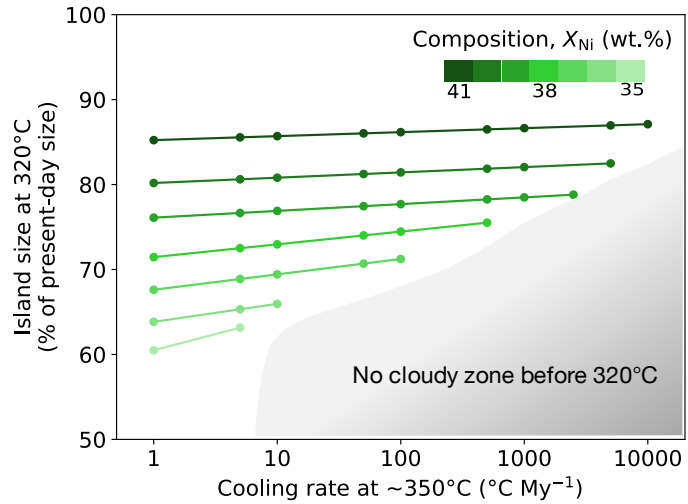

Fig. 5. Island size at 320°C (normalized by present-day island size) as a function of cooling rate for bulk composition of 35 to 41 wt.% Ni. The grey area encompasses conditions where the system does not form a relatively periodic pattern (i.e., cloudy zone) before reaching 320°C. Note that this does not prevent the CZ from forming below this temperature in this region. We speculate that islands should form directly with the tetrataenite structure in that case.

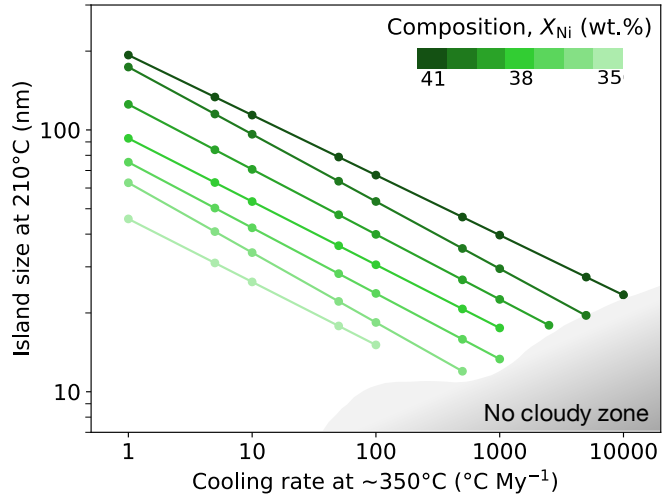

Fig. 6. Island size at 210°C (essentially equal to the present-day island size), as a function of cooling rate for bulk compositions between 35 and 41 wt.% Ni. The grey area shows where the cloudy zone does not form or where the compositions of islands and matrix do not have time to reach their expected final composition.

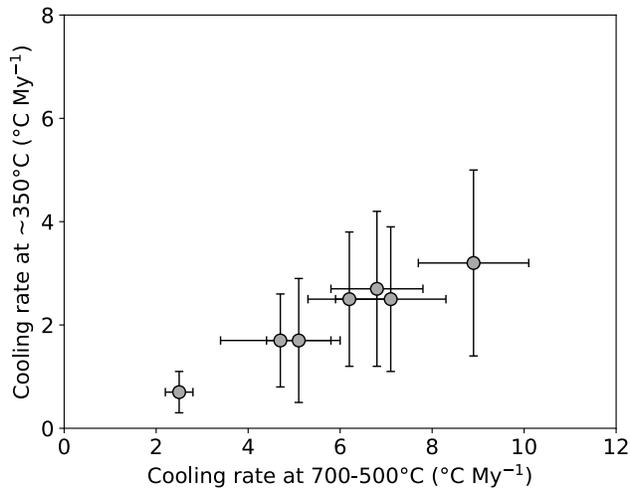

Fig. 7. Cooling rate at ~350°C for seven main-group pallasites as inferred from our model using a Ni content between 39 and 40 wt.% as a function of published cooling rate at 700-500°C (Yang et al., 2010). The average ratio of the cooling rate below 350°C and the cooling rate at 700-500°C is about 0.35.

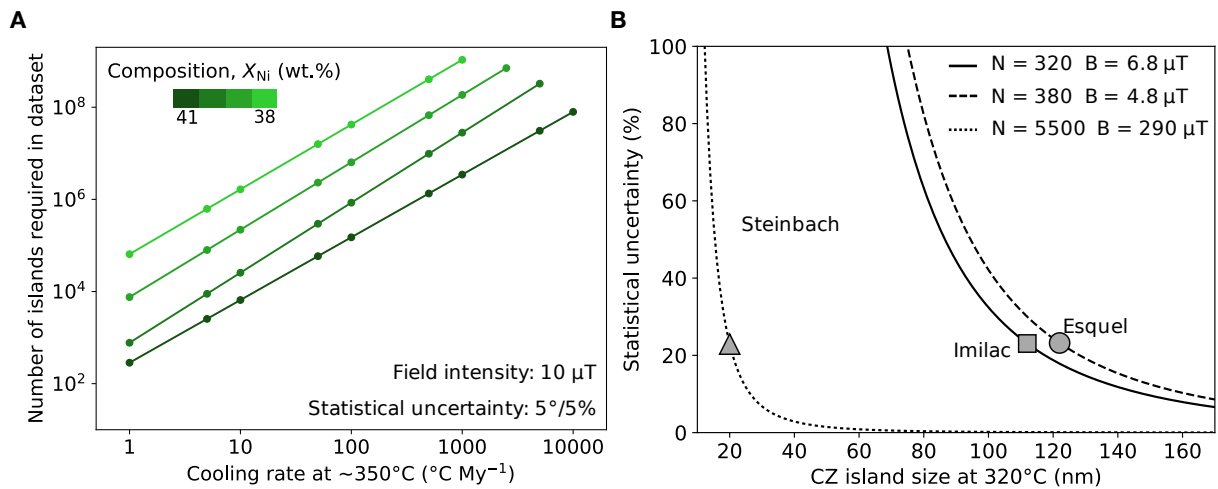

Fig. 8. A) Number of islands required per XPEEM dataset to limit the statistical uncertainty (due to the limited number of CZ islands) to 5° in paleodirection and 5% in paleointensity. This number is plotted as a function of cooling rate and local Ni content. It is obtained by combining island sizes at blocking temperature of 320°C provided by our model with the derivation of Berndt et al. (2016) for a Curie temperature of 550°C and assuming an ancient field of 10 µT. B) Statistical uncertainty in paleointensity as a function island size at blocking temperature (for angular statistical uncertainty, see Supplementary Figure S2.1). Lines represent different combinations of ancient field intensity and number of islands corresponding to previous XPEEM studies (Table 1). Markers show the island size at 320°C and associated statistical uncertainty for each meteorite studied.

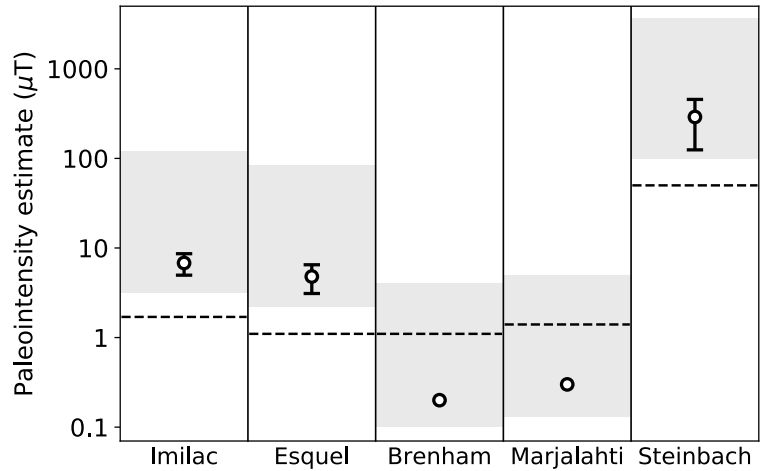

Fig. 9. Initial and improved paleointensity estimates from previously published XPEEM studies (Bryson et al. 2015, Nichols et al. 2016 and Bryson et al. 2017). The grey intervals show the initial range of paleointensities from the original publications: the upper bound is obtained with islands 30% of present-day size at 320°C, the lower bound is obtained for islands of present-day size at 320°C. The dotted lines show the simulated upper limit in intensity that would be measured with XPEEM if the meteorites had cooled in the absence of a field (see Section 5.1). Points show the improved paleointensity estimates using the island size at 320°C provided by our model. The error bars account for the $2\sigma$ measurement uncertainty, the $2\sigma$ statistical uncertainty and the $2\sigma$ uncertainty in island size at 320°C. Given that the mean paleointensities for Brenham and Marjalahti fall below the zero-field threshold, we cannot reject the hypothesis that these meteorites cooled through 320°C in the absence of a field. We did not include the error bars of these meteorites for clarity: because the statistical uncertainty is inversely proportional to the field intensity, the error bars would be very large but would not change the conclusion above.

| Meteorite | Group | Present-day island size (nm) | Assumed Ni content near the rim (wt.%) | Island size at 320°C (nm) | Predicted cooling rate below 350°C (°C My$^{-1}$) | Number of islands in XPEEM datasets | Statistical error in intensity (%) | Improved paleointensity estimates (µT) |
|---|---|---|---|---|---|---|---|---|
| Imilac | MG Pallasite | 143 ± 4 | 39 – 40 | 112 | 1.2 ± 0.7 | 320 | 23 | 6.8 ± 2.0 |
| Esquel | MG Pallasite | 157 ± 11 | 39 – 40 | 122 | 0.9 ± 0.5 | 380 | 23 | 4.8 ± 1.5 |
| Brenham | MG Pallasite | 123 ± 3 | 39 – 40 | 96 | 2.5 ± 1.4 | 1800 | – | 0.2 (< zero-field 1.1 µT) |
| Marjalahti | MG Pallasite | 118 ± 3 | 39 – 40 | 92 | 2.9 ± 1.5 | 1480 | – | 0.3 (< zero-field 1.4 µT) |
| Steinbach | IVA Iron | 29 ± 3 | 37 – 38 | 20 | 56.5 ± 25 | 5500 | 22 | 290 ± 165 |

Table 1. Meteorites previously analyzed by XPEEM. The first and second columns list the names and groups of the meteorites. The island sizes in the third column are from Yang et al. (2010) for Imilac, Esquel, Brenham and Marjalahti and Goldstein et al. (2009b) for Steinbach. The fourth column is the bulk Ni content near the tetrataenite rim assumed in our model to calculate an estimate of the cooling rates below ~350°C. The fifth column shows the island size at 320°C provided by the model. The sixth column lists the predicted cooling rates below ~350°C. The seventh column shows the number of islands included in each XPEEM dataset. The eighth column gives the statistical uncertainty for each XPEEM datasets. Finally, the ninth column summarizes the improved paleointensity estimates with their total uncertainty; note that these averages do not account for the possible effect of magnetostatic interactions between islands.

Bryson, J.F.J., Weiss, Scholl, A., Young, A.T. and Nimmo, F., Scholl, A. (2016) Paleomagnetic evidence for a partially differentiated H chondrite parent planetesimal, abstract 1546, 47$^{th}$ Lunar and Planetary Sci. Conf., The Woodlands, TX, March 21–25.

Bryson, J.F.J, Weiss, B.P., Harrison, R.J., Herrero-Albillos, J. and Kronast, F. (2017) Paleomagnetic evidence for dynamo activity driven by inward crystallisation of a metallic asteroid, Earth Planet. Sci. Lett. 472, 152–163.

Buchwald, V.F., 1975. Handbook of Iron Meteorites, Volume 1, University of California Press, New York.

Cacciamani, G., Dinsdale, A., Palumbo, M. and Pasturel, A. (2010) The Fe–Ni system: Thermodynamic modelling assisted by atomistic calculations, Intermetallics 18, 1148–1162.

Cahn, J.W. (1965) Phase separation by spinodal decomposition in isotropic systems, J. Chem. Phys. 42, 93–99.

Cahn, J.W. (1966) The later stages of spinodal decomposition and the beginnings of particle coarsening, Acta Metall. Mater. 14, 1685–1692.

Cahn, J.W. (1968) Spinodal decomposition, T. Metall. Soc. AIME 242, 166–180.